\documentstyle[prd,preprint,aps]{revtex}
\begin{document}

\reversemarginpar
\tighten

\title{Extensive Entropy Bounds}  
\author{Gilad Gour\thanks{E-mail:~gilgour@Phys.UAlberta.CA}}
\address{Theoretical Physics Institute, 
Department of Physics, University of Alberta,\\
Edmonton, Canada T6G 2J1}

\maketitle

\begin{abstract}
It is shown that, for systems in which the entropy is an extensive
function of the energy and volume, the Bekenstein and the holographic
entropy bounds predict new results. 
More explicitly, the Bekenstein entropy bound leads to the entropy
of thermal radiation (the Unruh-Wald bound) and the spherical
entropy bound implies the ``causal entropy bound". Surprisingly, the 
first bound shows a close relationship between black hole physics and the 
Stephan-Boltzmann law (for the energy and entropy flux densities
of the radiation emitted by a hot blackbody). Furthermore,
we find that the number of different species of massless fields is 
bounded by $\sim 10^{4}$.
   
\end{abstract}

\pacs{PACS numbers:~}

According to classical general relativity, ordinary matter that 
crosses the event horizon will disappear into the spacetime 
singularity of the black hole. In 1971, J.~A.~Wheeler raised the
question: what happens to the entropy
initially present in the matter? It seems that there is no gain of 
ordinary entropy in the universe, to compensate for the loss of 
entropy of the matter that has been absorbed by the black hole. Therefore,
the second law of thermodynamics is violated in this process. 

Bekenstein~\cite{Bek73} found a way out of this paradox by introducing the 
notion of black hole entropy. He assigned an entropy $S_{BH}$ that is 
proportional
to the horizon surface area of the black hole. After the discovery of Hawking
radiation, this entropy has been elevated to the status of a physical theory.
Furthermore, Bekenstein proposed to replace the ordinary second law of 
thermodynamics by the {\it Generalized Second Law} (GSL): 
The generalized entropy, $S_{g}\equiv S+S_{BH}$, of a system consisting 
of a black hole and ordinary matter (with entropy $S$) never decreases
with time (for an excellent review on the thermodynamics of black holes 
and the validity of the GSL, see Wald~\cite{W01}).

The GSL not only resolves the difficulty emphasized by Wheeler but also
imposes entropy bounds to hold for arbitrary systems. The first entropy 
bound was proposed
by Bekenstein~\cite{Bek81} more than 20 years ago. He considered a 
{\it gedanken-experiment} such that one lowers adiabatically 
a spherical box of radius $R$ toward a black hole (Geroch process). 
The box is lowered from infinity
where the total energy of the box plus matter contents is $E$.
It was shown~\cite{Bek81} that the entropy $S$ of the box must obey
(throughout the paper $c=k_{B}=1$)
\begin{equation}
S\le {2\pi R E\over\hbar},
\label{bb}
\end{equation} 
in order to preserve the GSL.  

The derivation of Eq.~(\ref{bb})  
in~\cite{Bek81} was criticized by Unruh and Wald~\cite{UW82,UW83} who
have argued that, since the process of lowering the box is a quasi-static
one (and therefore can be considered as a sequence of static-accelerating 
boxes), the box should experience a buoyant force due to the Unruh 
radiation~\cite{Unr76}.
Describing the acceleration radiation as a fluid, they have shown that this 
buoyant force alters the work done by
the box such that no entropy bound in the form of Eq.~(\ref{bb})
is necessary for the validity of the GSL. A few years ago, 
Pelath and Wald~\cite{PW} gave further arguments in favor    
of this result. 

Bekenstein~\cite{Bek83,Bek94}, on the other hand, argued that, only 
for very flat systems, the Unruh-Wald effect may be important. Later on, 
he has shown~\cite{Bek99} that, if the box is not almost at the horizon, the 
typical wavelengths in the radiation are larger than the size of the box
and, as a result, the derivation of the buoyant force from a fluid picture 
is incorrect.    
The question of whether the Bekenstein bound follows from the GSL via
the {\it Geroch process} remains controversial 
(see~\cite{And99,W01,Bek01,MS02}). 
However, as it was shown by Bousso~\cite{BouBek} (see the following paragraphs), 
there is another {\it link} connecting the GSL with the Bekenstein bound.

Susskind~\cite{Sus95} has shown, by considering the conversion of a 
system to a black hole, that the GSL implies a spherical entropy bound
\begin{equation}
S\le {1\over 4l_{p}^{2}}A,
\label{sb}
\end{equation}
where $S$ is the entropy of a system that can be enclosed by a sphere
with area $A$. A few years later, Bousso~\cite{Bou99,Bou02} had found 
an elegant way to generalize Eq.~(\ref{sb}) and write it in a covariant
form. He proposed the {\it covariant
entropy bound}: ``the entropy on any light-sheet $L(B)$ of a surface $B$
will not exceed the area of $B$". That is,
\begin{equation}
S[L(B)]\le\frac{A(B)}{4l_{p}^{2}},
\label{ceb}
\end{equation}
where the light-sheet $L[B]$ is constructed by the light rays that emanate
from the surface $B$ and are not expanding (for an excellent review
see Bousso~\cite{Bou02}). 

When a matter system with initial entropy $S$ falls into a black hole, the 
horizon surface area increases at least by $4l_{p}^{2} S$ due to the GSL.
This motivated Flanagan, Marolf and Wald~\cite{FMW} to generalize  
Eq.~(\ref{ceb}) into the following form:
\begin{equation}
S[L(B,B')]\le\frac{A(B)-A(B')}{4l_{p}^{2}},
\label{gceb}
\end{equation} 
where $L(B,B')$ is a light-sheet which starts at the cross-section
$B$ and cuts off at the cross-section $B'$ before it reaches a caustic.

Unlike the controversial issues regarding the relationship between the 
GSL and Eq.~(\ref{bb}), the entropy bounds in 
Eqs.~(\ref{sb},\ref{ceb},\ref{gceb}) are closely related to the GSL.
However, very recently, Bousso~\cite{Bou02} has shown that 
the Bekenstein entropy bound follows from Eq.~(\ref{gceb}) for any
isolated, weakly gravitating system. Hence,  
even though it is not clear whether quantum effects should be taken
into consideration in the derivation of Eq.~(\ref{bb}) (via the Geroch 
process), there is a strong link between the GSL and the Bekenstein 
bound. 

In the following, we provide another link connecting the 
bound~(\ref{bb}) with the entropy of thermal radiation and the 
Stephan-Boltzmann law. In our derivation, we consider systems in 
which the entropy density is a function of the energy density.
Later on, we show that for such systems, the spherical entropy 
bound~(\ref{sb}) yields the {\it causal entropy bound} proposed
by Brustein and Veneziano~\cite{BV00} and independently by 
Sasakura~\cite{Sasa}. We conclude that our results provide universal 
upper bounds for extensive systems.     
  
Consider an isolated spherical box\footnote{Throughout the paper we shall 
assume a spherical symmetry even though it is not always necessary.} 
of size $R$ and volume 
$V={4\pi\over 3}R^{3}$. Let us denote by $S(E,V)$ 
the {\it maximum} entropy of the box under the condition 
that $S$ is an extensive function of $E$ and $V$. 
Bekenstein's bound in terms of $E$ and $V$ is given by:
\begin{equation}
S(E,V) < \eta {EV^{1/3}\over \hbar},
\label{bbn}
\end{equation}
where $\eta =(6\pi ^{2})^{1/3}$.

Since the maximum entropy, S(E,V), will preserve the extensivity
property of entropy, it can be written as follows:
\begin{equation}
S(E,V)=VF\left(\frac{E}{V}\right)\;,
\label{Eul}
\end{equation}
where $F$ is some function of the energy density, 
$\varepsilon\equiv E/V$.
Eq.(\ref{Eul}) is equivalent to Euler's theorem on 
homogeneous functions\footnote{We assume the case where there are no other 
thermodynamic functions such as an electric or chemical potential.}:
\begin{equation} 
E{\partial S\over\partial E} +V{\partial S\over\partial V}=S\;.
\end{equation}

Now, Eq.(\ref{bb}) and Eq.(\ref{Eul}) imply a bound on F:
\begin{equation}
F(E/V) < {\eta E\over V^{2/3}\hbar}\;.
\end{equation}
In order to compare between dimensionless quantities,
let us multiple Eq.(3) by $l_{p}^{3}$ and define the following
dimensionless quantities:
\begin{equation} 
x \equiv {l_{p}^{3} E \over \epsilon _{p}V}\;,\;\; 
y \equiv {V^{1/3}\over l_{p}}\;,\;\;{\rm and}\;\; 
f(x) \equiv l_{p}^{3}F\left(\frac{E}{V}\right)\;,
\label{9}
\end{equation}
where $\epsilon _{p}$ and $l_{p}$ are the Planck energy and the 
Planck length, respectively.  
In these notations,
Eq.(3) can be written as
\begin{equation}
f(x) < \eta xy\;.   
\end{equation}

However, $x$ and $y$ can be considered as two independent 
parameters! Therefore, let us fix $x$ and take $y$ to its
minimal value. Reducing $y$ implies reducing {\it both} $R$ and $E$
since $x$ is kept constant (actually $E$ decreases faster than $R$). 
Hence, the minimal value of $y$ can be obtained by
the requirement $R > \hbar/ E$ (otherwise, the energy will leak out 
of the box). In terms of $x$ and $y$, it 
means $y > x^{-1/4}$. Thus, taking $y \sim x^{-1/4}$, we find
that
\begin{equation}
f(x) < x^{3/4},  
\end{equation}
where from this point we will stress functional dependence,
while ignoring numerical factors.

By substituting this condition in Eq.~(\ref{Eul}), we obtain the following 
``extensive entropy bound":
\begin{equation}
S(E,V) < {E^{3/4}V^{1/4}\over\hbar^{3/4}} 
= \left(\frac{ER}{\hbar}\right)^{3/4}\;.
\label{SB}
\end{equation}

The above result, by itself, is not surprising. For example,
consider a gas of radiation at temperature $T$ that is confined 
in the box. The energy and the entropy are given by the 
Stephan-Boltzmann law (neglecting corrections due to the discreteness 
of modes)
\begin{equation}
E\sim n_{s} R^{3}T^{4}\;\;\;{\rm and}\;\;\;S\sim n_{s} R^{3}T^{3},
\end{equation} 
where $n_{s}$ is the number of different (non-interacting) 
species of particles in the gas. Hence, in terms of $E$ and $R$, the 
entropy is proportional to 
\begin{equation}
S\sim n_{s}^{1/4}\left({ER\over\hbar}\right)^{3/4}. 
\label{species}
\end{equation}
That is, the entropy of thermal radiation
saturates Eq.~(\ref{SB}). It is a good guess that no other
system has more entropy, because the rest mass of 
ordinary particles only enhances gravitational instability without
contributing to the entropy. Thus, the bound~(\ref{SB}) is 
understandable. 

However, there are three
points about Eq.(\ref{SB}) that are very interesting and somewhat 
surprising:
First, one does {\it not} have to define unconstrained thermal radiation 
to be the maximum entropy system (as did, for example, Unruh and 
Wald~\cite{UW82,UW83} 
and Pelath and Wald~\cite{PW}). It comes out that the entropy of extensive 
systems is no higher if one assumes Bekenstein's bound.

Second, the GSL leads to Bekenstein's bound and extensivity 
leads to a bound proportional to the thermal radiation entropy! 
That is, the GSL implies the Stephan-Boltzmann Law.
Boltzmann and other physicists in his time would have 
never imagined that one would be able to obtain the thermal radiation 
entropy from black hole physics\footnote{In some way, it also gives a 
further evidence of Bekenstein's identification of black hole entropy 
with the horizon area}. Let us take this moment to mention 
that both the black hole entropy formula, $S\propto A$
($A$ is the horizon surface area),
and the Stephan-Boltzmann formula, $u\propto T^{3}$ 
and $s\propto T^{4}$ ($u$ and $s$ are the energy and entropy flux densities
of the radiation emitted by a hot blackbody),  
can be derived purely by classical thermodynamics~\cite{GM}. 
This shows another similarity between the physics of black holes and 
blackbodies.  

Third, the species problem: one of the objections to {\it all} kinds
of entropy bounds is that one can take $n_{s}$ in Eq.(\ref{species}) 
to be arbitrarily large. Consider for example the spherical entropy bound.
In order for $S$ in Eq.(\ref{species}) to become greater than 
$A/4l_{p}^{2}$, one has to take $n_{s}> A/l_{p}^{2}$.
Of course, we have no evidence
(experimental or string theoretical) that $n_{s}$ can run into such a high
number, as would
be required to violate the bound. That is, one can always hold the
position that the bound is telling us about the world as it is, not as it
might be in the imagination of a physicist who needs counterexamples.
Furthermore, if the number of species grows, one can raise the question 
whether interactions will not nullify the assumption of ``free particles''.

However, the species problem manifests in a much more conspicuous form in 
the extensive bound~(\ref{SB}). This bound implies that $n_{s}^{1/4}$
must be of order unity. That is, $n_{s}$ can not be much greater than
$10^{4}$. This number is much smaller than $A/l_{p}^{2}$ and it raises
the question whether there are more realistic bounds on the number of
species in nature.
Since the arguments that lead to Eq.~(\ref{SB}) include the 
assumption that the minimum value of $R$ is approximately the Compton wave 
length $\hbar/E$, we could not obtain the exact dimensionless numerical factor 
that should be added to Eq.~(\ref{SB}). This numerical factor would have
provided an exact bound on $n_{s}$.   

In the above considerations, it was assumed that the system does not 
exceed Bekenstein's entropy bound. Let us now instead consider the 
the relationship between the spherical entropy bound~(\ref{sb})
and the extensivity property of the entropy function $S(E,V)$.
As we will see in the following, this relationship yields the ``causal
entropy bound"~\cite{BV00} which scales as $\sqrt{EV}$. We will first 
obtain this result by a simple heuristic argument, and then we will
prove it rigorously.

Consider a box of size $R$ (volume $R^{3}$) with
energy $E$. According to the Holographic bound~(\ref{sb}),
the entropy of the box can not exceed $\sim R^{2}/l_{p}^{2}$.
Now, consider $N^{3}$ ($N$ is an integer) identical 
boxes arranged in a much bigger box of size $NR$ 
(volume $N^{3}R^{3}$). If the interactions between the boxes
are negligible and $N$ is not too big (i.e. the big box is not 
a black hole), the entropy of the big box
is $N^{3}S$, where $S$ is the entropy of a single box.
However, by applying the Holographic bound for the big box,
we get: 
\begin{equation}
N^{3}S < {(NR)^{2}\over l_{p}^{2}} 
\end{equation}
or equivalently 
\begin{equation}
S < {R^{2}\over Nl_{p}^{2}}\;. 
\label{ent1}
\end{equation}
As it was expected, the Holographic bound implies a tighter bound 
for ``non-gravitating" systems. 

Now, the total mass (energy)
of the big box, $N^{3} E$, should be smaller than the size
of the box $NR$ (the big box is not a black hole). Therefore, 
the maximum possible value of $N$ is of the order 
$\sim (\epsilon _{p}R/l_{p}E)^{1/2}$. 
By supplementing this in Eq.~(\ref{ent1}) we get:
\begin{equation}
S \lesssim 
{R^{2}\over \hbar} \times\left(\frac{l_{p}E}
{\epsilon _{p}R}\right)^{1/2}\;\; \propto \; \sqrt{EV}\;,  
\label{ent2}
\end{equation}
where $V$ is the volume of the small box. 

The above arguments clarify why for extensive systems
the holographic principle predicts the bound~(\ref{ent2}). 
We shall now prove this bound in a more formal way:

{\it Theorem}: Denote by $S(E,V)$ the maximum entropy that 
an isolated spherical system with energy $E$ and volume $V$ can 
have, under the condition that the entropy is distributed uniformly.
The spherical entropy bound then implies that 
$S(E,V)\lesssim\sqrt{EV}$.

The entropy is distributed uniformly if and only if it can be
written in the form given in Eq.~(\ref{Eul}).  
On the other hand, 
the spherical entropy bound implies 
\begin{equation}
S(E,V) < {V^{2/3}\over\l_{p}^{2}}\; . 
\end{equation}
Therefore, the bound~(\ref{sb}) leads to the result:
\begin{equation}
F\left(\frac{E}{V}\right) < \frac{1}{l_{p}^{2}V^{1/3}}\;,
\end{equation}
where $F$ is defined in Eq.~(\ref{Eul}). In terms of the dimensionless
quantities which are defined in Eq.~(\ref{9}), the above inequality
can be written in the form
\begin{equation} 
f(x) < y^{-1}.
\end{equation}
However, the two dimensionless parameters 
$x$ and 
$y^{-1}$ can be considered as independent. Therefore, 
one can keep $x$ constant and take
$y^{-1}$ to its minimum value. The minimum value of $y^{-1}$ occurs 
when the size of the system $V^{1/3}$ becomes comparable with
its energy. That is, when $x = y^{-2}$. Hence,
\begin{equation}
f(x) < \sqrt{x}\;.
\label{den1}
\end{equation}
Eq.~(\ref{den1}) provides a proof for the theorem above. It shows a close
relationship between the holographic principle and the causal entropy bound
obtained by Brustein and Veneziano~\cite{BV00} and independently by 
Sasakura~\cite{Sasa}. 
In~\cite{BV00} the causal entropy bound is defined covariantly
and, hence, it is much more general then our derivation. Note, however, that
for weakly gravitating systems, the Bekenstein bound is tighter than
the causal bound so that the later may be useful only for strongly
gravitating systems. It may seem that the causal bound does not imply 
that the number of fundamental degrees of freedom is related to the area surfaces 
in spacetime. However, from our derivation of the causal bound (based on the 
spherical bound), we learn that the causal bound can be incorporated into 
a holographic world.   

In conclusion, in this paper we considered the applications of
the entropy bounds~(\ref{bb},\ref{sb}) into extensive systems.
It was shown that extensivity provides links
between different entropy bounds. One of the main results was the derivation
of a bound proportional to the entropy of thermal radiation from black hole physics.
In the future, we hope to generalize the results to charged and rotating systems.  
  
\section{Acknowledgments} 
I would like to thank J.~Bekenstein for his influence and inspiration.
I would also like to thank D.~Page and V.~Frolov for helpful discussions 
and to A.J.M.~Medved for reading and improving the English in the manuscript. 
The author is also grateful to the Killam Trust for its financial support.

\end{document}